%Paper: hep-th/9403054
%From: "Eva Silverstein" <silver@puhep1.Princeton.EDU>
%Date: Tue, 8 Mar 94 14:10:15 -0500

\input harvmac.tex
\def\bar{\overline}
\def\bq{{\overline Q_{+}}}
\def\bd{{\overline D_{+}}}
%\lref\calletal{C.G. Callan, D. Friedan, E.J. Martinec and M.J. Perry,
%``Strings in Background Fields'', {\it Nucl. Phys.}~{\bf B262} (1985)593;
%E.S. Fradkin and A.A. Tseytlin, ``Quantum String Theory Effective Action'',
%{\it Nucl. Phys.}~{\bf B261}~(1985)1. }
\lref\phases{E. Witten, ``Phases of N=2 Theories in Two Dimensions,''
{\it Nucl. Phys.} {\bf B403} (1993) 159,
hepth/9301042.}
\lref\min{E. Witten,``On the Landau-Ginzburg Description of N=2
Minimal Models'', IASSNS-HEP-93/10, hepth/9304026.}
\lref\spec{S. Kachru and E. Witten, ``Computing the Complete Massless
Spectrum of a Landau-Ginzburg Orbifold'', {\it Nucl. Phys.}
{\bf B407} (1993) 637, hepth/ 9307038.}
\lref\shamjac{J. Distler and S. Kachru, ``(0,2) Landau-Ginzburg Theory'',
Princeton Preprint PUPT--1419, hepth/9309110.}
%\lref\vafaetc{old N=2 stuff}
\lref\cpn{E. Witten, ``Instantons, the Quark Model, and the 1/N
Expansion'', {\it Nucl. Phys.} {\bf B149} (1979) 285;
A. D'Adda, P. Di Vecchia and M. Luscher, ``A 1/n Expandable
Series of Non-linear Sigma Models with Instantons'', {\it Nucl. Phys.}
{\bf B146} (1978)63}
\lref\fre{P. Fr\'e, L. Girardello, A. Lerda, and D. Soriani,
``Topological First-Order Systems with Landau-Ginzburg Interactions'',
{\it Nucl. Phys.}~{\bf B387} (1992)~333.}

\Title{IASSNS-HEP-94/4, PUPT-1453
}
{\vbox{\centerline{Global U(1) R-symmetry and}
	\vskip2pt\centerline{Conformal Invariance of (0,2) Models}}}
\centerline{Eva Silverstein\footnote{$^\dagger$}
{silver@puhep1.princeton.edu}}
\bigskip\centerline{Joseph Henry Laboratories}
\centerline{Jadwin Hall}
\centerline{Princeton University}\centerline{Princeton, NJ 08544 USA}
\bigskip\centerline{and}
\bigskip\centerline{Edward Witten\footnote{$^*$}
{witten@iassns.bitnet}}
\bigskip\centerline{School of Natural Sciences}
\centerline{Institute for Advanced Study}
\centerline{Olden Lane}
\centerline{Princeton, N.J. 08540}
\vskip .3in

%%ABSTRACT GOES HERE
We derive a condition under which (0,2) linear sigma models
possess a ``left-moving'' conformal stress tensor in $\bq$ cohomology
(i.e. which leaves invariant the ``right-moving'' ground states)
even away from their critical points.  At the classical level
this enforces quasihomogeneity of the superpotential terms.  The
persistence of this structure at the quantum level on the worldsheet
is obstructed by an anomaly unless the charges and superpotential
degrees satisfy a condition which is equivalent to the condition
for the cancellation
of the anomaly in a particular ``right-moving'' U(1) R-symmetry.

\Date{03/94} %replace this line by \draft  for preliminary versions
	     %or specify \draftmode at some point
%\draft

\newsec{Introduction}

The perturbative conditions for conformal
invariance of nonlinear sigma models have long
been studied.  Of particular relevance for string theory are N=2
superconformal theories.
Recently it has been noticed (see \phases~and \shamjac~ and
references therein) that (0,2) string
vacua can be studied
by computing renormalization group invariant
quantities in simpler (0,2) models that are not conformally invariant but
may flow in the infrared to conformal fixed points.
The models that are useful in this respect are linear sigma models,
in general coupled to gauge fields.

If one is to study (0,2) string vacua by studying linear sigma models
in the same universality class, one would like criteria for knowing
which linear sigma models are likely to have conformal fixed points in the
infrared.  For instance, consider the $CP^n$ model -- a well-known
(2,2) model which we will think of as a (0,2) model, ignoring the
left-moving supersymmetry.  One can certainly find a simple linear
sigma model with gauge fields that is equivalent to the $CP^n$ model
in the infrared.  One would not expect this model to flow in the infrared
to a non-trivial conformal field theory.
We would like to
understand what restrictions on the
gauge charges and superpotential interactions in linear sigma
models are necessary to exclude models such as the $CP^n$ model
which do not describe superstring vacua.

The global right-moving supercharges of a (0,2) or (2,2) model are operators
$\bq$ and $Q_+$ that obey $\bq^2=Q_+{}^2=0$,
$\{\bq,Q_+\}=P_+$; in particular, if we think of $\bq$ as a cohomology
operator, then the cohomology is the space of right-moving ground
states (states of $P_+=0$).
One of the main observations of \min~ was that at the
classical level a (2,2)
Landau-Ginzburg theory (with a quasi-homogeneous superpotential)
has a left-moving superconformal algebra even away from criticality.
Indeed, one can find operators commuting with $\bq$, and generating
by operator products the left-moving $N=2$ algebra
(modulo terms of the form $\{\bq,\dots\}$) directly in the non-critical
Landau-Ginzburg theory.  It is plausible to expect that the $\bq$-trivial
error terms
$\{\bq,\dots\}$ vanish in the infrared limit and to interpret the left-moving
$N=2$
algebra that one finds away from criticality as a precursor
of the conjectured superconformal fixed point.

We want to carry out here a similar program for (0,2) models.
A model possessing
a (0,2) superconformal fixed point will at that fixed point
have
a left-moving stress-tensor $T_{--}$
which satisfies the conformal algebra and commutes with the right-moving
global supersymmetry charges.  We will determine which linear models
possess such a left-moving conformal symmetry away from criticality
at the level of $\bar Q_+$
cohomology.  In the process, we will also generalize some of the previous
results concerning (2,2) models.

Finding a left-moving conformal symmetry at the level of $\bar Q_+$
cohomology in the non-critical model is neither necessary nor sufficient
to ensure the existence of a conformally invariant infrared fixed point.
It is obviously not sufficient, a priori.
It is also not necessary in general,
since if a conformal limit exists, the left-moving
stress tensor that commutes with $\bar Q_+$ might appear only at the fixed
point.  (The stress tensor would have to appear paired
with another new state in the $\bar Q_+$ cohomology with right-moving $U(1)$
charge differing by one and the same left-moving quantum numbers.
In many instances the $\bar Q_+$ cohomology of the conformal theory
has no suitable states; in that case the existence of the left-moving
conformal symmetry mod $\{\bar Q_+\}$ in the non-critical
theory is indeed necessary.)
Nevertheless, we think that the occurrence of the off-shell
conformal symmetry is a very interesting hint of the existence of
the fixed point.

In section 2 we will explain the classical condition that arises from
this requirement.  The classical condition is not enough: even if
the left-moving conformal symmetry arises classically, there may be
a quantum anomaly.
In section 3 we compute
quantum anomaly in
[$\bq$, $T_{--}$] and relate this to the anomaly in a particular
right-moving R-symmetry.

\newsec{The Classical Condition}

We follow the conventions in
\phases~ so that our action is
\eqn\action{\eqalign{S=
&\int d^2y d^2\theta \bigl\{{1\over{8e^2}}Tr\bar\Upsilon\Upsilon
-{i\over{2}} \Phi_i(D_0-D_1)\bar\Phi_i -
{1\over{2}}\bar\Lambda_{-a}\Lambda_{-a}\bigr\}\cr
&+(\int d^2y d\theta^+\bigl\{ {t\over{4}}\Upsilon\bigl|_{\bar\theta=0}
-{1\over{\sqrt{2}}}\Lambda_{-a}J^a\bigl|_{\bar\theta=0}\bigr\}+h.c.)\cr}}
Here we have
$N_i$ chiral multiplets $\Phi_i=\phi_i+{\sqrt{2}\theta^+\psi_{i+}}
+{i\theta^+\bar\theta^+(D_0+D_1)\phi_i}$ with gauge charges $Q_i$,
$N_a$ fermionic multiplets
$\Lambda_a=\lambda_{-a}-{\sqrt{2}\theta^+G_a}-{i\theta^+\bar\theta^+
(D_0+D_1)\lambda_{-a}}-{\sqrt{2}\bar\theta^+E_a}$
with gauge charges $Q_a$,
and a U(1) gauge multiplet with fermionic field strength
$\Upsilon=-\chi_-+{\theta^+(v_{01}+iD)}+{i\theta^+\bar\theta^+
(D_0+D_1)\chi_-}$.  $E_a$ and $J^a$ are holomorphic functions
of the $\Phi_i$.
The fermionic multiplets satisfy $\bd\Lambda_{-a} = \sqrt{2}E_a(\Phi_i)$,
and $E_aJ^a =0$.  Here all derivatives are gauge-covariant derivatives.

Consider the following candidate stress tensor (a slight generalization
of that in \fre~ and \min):
\eqn\stress{
\eqalign{
T_{--}
&= \biggl( ({-i\over {2e^{2}}})
\Upsilon (D_0-D_1) \bar\Upsilon +
2\,(D_0-D_1)\Phi_i(D_0-D_1)\bar\Phi_i
\cr
&+i[\Lambda_{-a}(D_0-D_1)\bar\Lambda_{-a}-
(D_0-D_1)\Lambda_{-a}\bar\Lambda_{-a}]
\biggr)
\cr
& - \sum_i\alpha_i(D_0-D_1)[\Phi_i(D_0-D_1)\bar\Phi_i]
+i \sum_a\alpha_a (D_0-D_1)(\Lambda_{-a}\bar\Lambda_{-a})
\biggl|_{\theta=\bar\theta=0}
}}
As discussed in \min, one can study $\bq$ cohomology by
studying $\bd$ cohomology since the two operators are
conjugate:
$$\bq=\exp\bigl[-2i\theta^+\bar\theta^+\bigr]\bd
\exp\bigl[2i\bar\theta^+\theta^+\bigr]$$
Using the equations of motion
\eqn\eomg{{1\over{4e^2}}\partial_-\bd\bar\Upsilon
= iQ_i\Phi_i(D_0-D_1)\bar\Phi_i + Q_a\bar\Lambda_{-a}\Lambda_{-a}}
\eqn\eomch{\bd(D_0-D_1)\bar\Phi =
-i\sqrt{2}(\Lambda_{-a}{\partial J_a\over{\partial\Phi_i}}-
\bar\Lambda_{-a}{\partial E_a\over{\partial\Phi_i}})}
\eqn\eomferm{\bd\bar\Lambda_{-a}=-\sqrt{2}J_a}
and the constraint
\eqn\constraint{\bd\Lambda_{-a} = \sqrt{2}E_a}
we find that
$\bd T_{--} =0$ classically provided that the
following quasihomogeneity conditions on $E_a$ and $J_a$ are satisfied:
\eqn\quasJ{\alpha_aJ^a + \sum_i {\alpha_i\Phi_i{{\partial J^a}
\over{\partial\Phi_i}}} = J^a}
and
\eqn\quasE{-\alpha_aE_a + \sum_i{\alpha_i\Phi_i{{\partial E_a}
\over{\partial\Phi_i}}} = E_a}
These conditions reduce to the usual quasihomogeneity condition
on the superpotential in the (2,2) case.  $T_{--}$ is not
$\bd$(...), so $T_{--}$ is a nontrivial element of $\bd$
cohomology.

We would like to understand the operator algebra satisfied by $T_{--}$ as
an element of the chiral algebra of the (0,2) model, i.e.
in $\bd$ cohomology.  One finds that the superpotential interactions
do not contribute to the singularities in the OPE. To see
this, note that since $\phi$ has mass dimension zero, the
superpotential terms $|E_a|^2+|J_a|^2$ must come with a
dimensionful coupling constant $\mu^2$.  Then (with $x^2=x_+x_-$)
the first superpotential
corrections to the OPE $T_{--}(x)T_{--}(0)$
would be of the
form ${\mu^2 x^2\over{x_-^4}}$, ${\mu^2x^2\over{x_-^2}}O_{--}$,
and ${\mu^2x^2\over{x_-}}\partial_-O_{--}^{\prime}$, for some operators
$O_{--}$ and $O_{--}^{\prime}$,
and so would vanish as $x_+\rightarrow 0$;
i.e. for Lorentz invariance $\mu^2$ would always be accompanied by
$x^2$.

%Another way to see this is to note that \spec
%\eqn\ham{\eqalign{L_{0-}
%&=\int  dx^1\bigl(D_-\phi_iD_-\bar\phi + {i\over{2}}
%\lambda_{-a}D^{\leftrightarrow}_-\bar\lambda_{-a}\cr
%&-{\alpha_i\over{2}}\partial_-(\phi_iD_-\bar\phi_i)
%+{i\alpha_a\over{2}}\partial_-(\lambda_{-a}\bar\lambda_{-a})\cr
%&+{i\over{8e^2}}\chi_-\partial^{\leftrightarrow}_-\bar\chi
%+{i\over{8e^2}}\partial_-(\chi_-\bar\chi_-)\bigr)\cr
%& + \{\bq,...\}\cr}}
%In particular note that the stress tensor \stress~ gives the
%first term of \ham.  The corresponding Lagrangian is, mod
%$\{\bq,...\}$, just the free-field Lagrangian plus total
%derivative terms.

Therefore we can compute the OPE~
$T_{--}(x)T_{--}(y)$  using the free propagators
$$<\bar\phi (x)\phi(y)> = log(x-y)^2,$$
$$<v_{\mu}(x)v_{\nu}(y)> = e^2 \eta_{\mu\nu}log(x-y)^2,$$
and
$$<\bar\lambda_{-a} (x) \lambda_{-a} (y)> = {1 \over{(x_- -y_-)}}
=<\bar\chi_-(x)\chi_-(y)>{1\over{4e^2}}.$$

We find that this stress tensor satisfies the conformal algebra
\eqn\ope{T_{--}(x)T_{--}(y)\sim {c/2\over{(x_--y_-)^4}}
+{2T_{--}(y)\over{(x_--y_-)^2}}
+{\partial T_{--}(y)\over{x_--y_-}}}
with
\eqn\centch{c=3\sum_i(1-2\alpha_i) + (N_a-N_i) + \sum_i3{\alpha_i}^2
-\sum_a3{\alpha_a}^2-\sum_g2}
where the last sum is over the generators of the gauge group.

Note that if we shift the $\alpha_i$ by $Q_i$ and $\alpha_a$
by $Q_a$, the quasihomogeneity conditions are still satisfied
by virtue of gauge invariance but the central charge shifts by
an amount proportional to $\sum_i{\alpha_i Q_i} - \sum_a{\alpha_a Q_a}
-\sum_i{Q_i}$, appearing to yield a family of conformal stress
tensors.  We will see in the next section that this situation will
be avoided by the quantum anomaly.

\newsec{The Anomaly in [$\bq$,$T_{--}$]}

In order to determine whether we can maintain [$\bq$,$T_{--}$]=0
at the wordsheet ``quantum'' level,
we compute the following time-ordered product:
\eqn\thrpt{0=\int{d^2x}\partial_{\mu}{T(S^\mu(x)T_{--}(y)
\bar W(z))}
= T([\bq,T_{--}(y)]\bar W(z))
+ T(\{\bq,\bar W(z)\} T_{--}(y))}
where $S^\mu$ is the supersymmetry current whose charge is $\bq$.
For this formula to be useful, $\bar W$ should be an operator whose
commutator with $\bq$ is known; in that case, the above formula
gives information about $[\bq,T_{--}]$.  To ensure that the commutator
of $\bar W$ with $\bq$ is known, we will take $\bar W$ to be one of the
elementary fermion fields of the model.

%Check convention:
If we take $\bar W_1 = \partial_-\bar\Upsilon\bigl|_{\theta=0=\bar\theta}
=(-2i)\partial_-\bar\chi_-$ then using the equation of motion
\eqn\eom{{1\over{4e^2}}\partial_-\bd\bar\Upsilon =
iQ_i\Phi_i(D_0-D_1)\bar\Phi_i + Q_a\bar\Lambda_{-a}\Lambda_{-a}}
we obtain
\eqn\varw{\{\bq,\bar W_1\} = (-2i)(4e^2)\bigl[iQ_i\phi_i(D_0-D_1)
\bar\phi_i + Q_a\bar\lambda_{-a}\lambda_{-a}\bigr]}
For the most singular terms in $T(\{\bq,\bar W_1(z)\} T_{--}(y))$
we can again use the free propagators since the only possible
contribution from the superpotential terms would be proportional
to ${\mu^2 x^2\over{x_-^3}}$, which would vanish as
$x_+\rightarrow 0$.
Computing using the free propagators, \stress,~ and \varw, we have
the most singular terms
\eqn\preanom{\eqalign {{1\over{(2i)(4e^2)}}
T(\{\bq,\bar W_1(z)\} T_{--}(y))\sim &
T(2\partial_-\phi_i\partial_-\bar\phi_i(y)
{}~iQ_i\phi_i\partial_-\bar\phi_i(z))\cr
&-T(\alpha_i\partial_-(\phi_i\partial_-\bar\phi)(y)~
iQ_i\phi_i\partial_-\bar\phi_i (z))\cr
&+T(i\alpha_a\partial_-(\lambda_{-a}\bar\lambda_{-a})(y)~
Q_a\lambda_{-a}\bar\lambda_{-a} (z))\cr
&\sim {{2i}\over{(y_- - z_-)^3}}(\sum_iQ_i - \sum_iQ_i\alpha_i
+\sum_aQ_a\alpha_a)\cr
&{\rm + less~ singular}\cr}}
from which we deduce
\eqn\anomaly{[\bq,T_{--}]=2i\partial_-\chi_-(\sum_iQ_i - \sum_iQ_i\alpha_i
+\sum_aQ_a\alpha_a)}
plus possibly other terms whose OPEs with
$\bar W_1=(2i)\partial_-\chi_-$ are nonsingular.
%Other choices for $\bar W$ lead to the classical quasihomogeneity
%conditions \quasJ~ and \quasE.
Considering other $\bar W$'s linear
in the fundamental fields yields no further divergent contributions
to $T(\{\bq,\bar W\}T_{--})$ at the one-loop level.
The contribution \preanom~
is proportional to $q_-^3/q^2$ in momentum
space (where $q$ is the momentum carried by $T_{--}$)
and cannot be cancelled by any local counterterm.
We have found that the existence of $T_{--}$ requires
\eqn\condition{\sum_iQ_i - \sum_iQ_i\alpha_i
+\sum_aQ_a\alpha_a = 0}

This condition \condition~ for the cancellation of this anomaly in
$[\bq,T_{--}]$
is the same as the condition for the cancellation of the anomaly
in the following right-moving R-symmetry:
\eqn\rsymm{\eqalign{
&\psi_{+i}\rightarrow e^{i\epsilon(1-\alpha_i)}\psi_{+i}\cr
& \chi_-\rightarrow e^{-i\epsilon}\chi_-\cr
&\lambda_{-a}\rightarrow e^{-i\epsilon\alpha_a}\lambda_{-a}\cr
&\phi_i\rightarrow e^{-i\epsilon\alpha_i}\phi_i\cr}}
Note that it is not sufficient to have ${\it some}$ right-moving
R-symmetry; we need this particular one.  The R-symmetry that arises
in this way has the following property: in case of a (0,2) linear
sigma model that happens to be a (2,2) model, it commutes with
the left-moving supersymmetry.

The above condition rules out the $CP^n$ model.  At the classical
level, the $CP^n$ model is a (2,2) model with left- and right-moving
R symmetry (and can be derived from a linear sigma model with those
properties).  At the quantum level, the axial R symmetry is anomalous
but the vector symmetry survives; let us call it $V$.
If one ignores the left-moving
supersymmetry, then $V$ transforms the right-moving supersymmetries
as an R symmetry, so one could view the $CP^n$ model as a (0,2) model
with a right-moving R symmetry.  However, the condition for left-moving
conformal invariance is not merely that there should be an anomaly-free
right-moving R symmetry, but that the particular symmetry in
\rsymm\ should be anomaly-free; this is not so for the $CP^n$ model.

In the (2,2) case the above  condition for cancellation of the anomaly
reduces to the condition
$\sum_iQ_i = 0$, which together with gauge invariance
was shown in \phases~ to reproduce the Calabi-Yau condition
in the appropriate limit (r=Re(t)$\gg$0).

Distler and Kachru recently analyzed the string theories arising
from (0,2) linear sigma models
in their Landau-Ginzburg phases \shamjac.  They not only
imposed the condition derived here for the right-moving R-symmetry
but also insisted on a non-anomalous left-moving U(1) current J.
In view of \shamjac~ one might wonder about the condition for the chirality
of J, which fills out the spacetime gauge group
and plays the important role of defining the GSO projection and
orbifold twisting.

For
\eqn\Jleft{J = (1-\tilde\alpha_a)\Lambda_{-a}\bar\Lambda_{-a}
-i\tilde\alpha_i\Phi_i D_-\bar\Phi_i}
a similar calculation to the one presented above
reveals that $[\bq,J]$ vanishes at the classical level given
the quasihomogeneity conditions
\eqn\quasJtwo{\tilde\alpha_aJ^a +
\sum_i {\tilde\alpha_i\Phi_i{{\partial J^a}
\over{\partial\Phi_i}}} = J^a}
and
\eqn\quasEtwo{-\tilde\alpha_aE_a +
\sum_i{\tilde\alpha_i\Phi_i{{\partial E_a}
\over{\partial\Phi_i}}} = -E_a}
That is, in order to impose the existence of the left U(1) as
well as the left stress tensor,
one needs to find constants $\alpha_a$, $\alpha_i$,
$\tilde\alpha_a$ and $\tilde\alpha_i$
and polynomials $E_a$ and $J_a$ to satisfy \quasJ, \quasE,
\quasJtwo, and \quasEtwo.  The cases considered in \shamjac~
had $E_a=0$ and satisfied these conditions with
$\tilde\alpha_a=\alpha_a$ and $\tilde\alpha_i=\alpha_i$.
In the gauged (2,2) case discussed in \phases~ one satisfies
these conditions with $\tilde\alpha_{\Sigma}=-\alpha_\Sigma$
for the gauge field strength $\Sigma$
and for all other fields $\tilde\alpha=\alpha$.
At the quantum level we obtain the condition
$\sum_a{Q_a}-\sum_a{\tilde\alpha_a Q_a} + \sum_i{\tilde\alpha_i Q_i} = 0$.
This is just the standard condition for the symmetry generated
by $J$ to be free of gauge anomalies -- in contrast to the left-moving
conformal symmetry where the anomaly in $\{\bq,\dots\}$ gave
new information.
\vskip .5in
\centerline{{\bf Acknowledgements}}
We would like to thank J. Distler for
suggesting that we study the left U(1).
E.S. is supported by an N.S.F. Graduate Fellowship and an AT$\&$T
GRPW Grant.  E.W. is supported in part by N.S.F. Grant
PHY92-45317.

\listrefs

\end